\documentclass[twocolumn,apj]{openjournal}

\usepackage[breaklinks,colorlinks,citecolor=blue]{hyperref}

\usepackage{natbib}
\usepackage{graphicx}

\usepackage[utf8]{inputenc}

\usepackage{amsfonts}
\usepackage{amssymb}

\usepackage{tikz}

\newcommand{\be}{\begin{equation}}
\newcommand{\ee}{\end{equation}}
\newcommand{\bea}{\begin{eqnarray}}
\newcommand{\eea}{\end{eqnarray}}
\newcommand{\Dd}{\mathrm{d}}
\newcommand{\pa}{\partial}
\newcommand{\fvec}[1]{\underline{#1}}

\begin{document}

\journalinfo{The Open Journal of Astrophysics}
\submitted{submitted 4 July 2020; accepted 28 July 2020}

\shorttitle{Cosmic event horizons and the light-speed limit for relative radial motion}
\shortauthors{M P\"ossel}

\title[Cosmic event horizons and the light-speed limit for relative radial motion]{Cosmic event horizons and the\\ light-speed limit for relative radial motion}
% In a long title you can use \\ to force a line break at a certain location.

\author{Markus P{\"o}ssel$^{\star1}$}
\address{$^1$Haus der Astronomie and Max Planck Institute for Astronomy, K{\"o}nigstuhl 17, 69124 Heidelberg, Germany}
%\ead{poessel@hda-hd.de}
%\author{[author name anonymized]}
%\address{[institution and address anonymized]}
%\email{poe}

\thanks{$^\star$ E-mail: \nolinkurl{poessel@hda-hd.de}}

\begin{abstract}
Cosmic event horizons separate spacetime into disjoint regions: those regions whose light signals can reach us, and more distant regions we cannot, even in principle, observe. For one type of cosmic horizon, associated with universes that keep expanding forever, there is a simple intuitive picture of where the cosmic horizon is located: As our own speed relative to a distant galaxy approaches the speed of light, it will become increasingly difficult for light from that galaxy to catch up with us. Light from galaxies in the limiting case where the relative speed reaches the speed of light should not be able to catch up with us at all; such galaxies and their more distant kin are beyond our cosmic horizon. Applied to the usual recession speeds of galaxies, that simple picture is wrong. But there exists an alternative definition of relative speed, derived from the relativistic concept of transporting four-velocities from one event to another for comparison. This article shows how, using that alternative concept of relative velocity, key elements of the intuitive catching-up picture are valid, and can be used in a simplified explanation for cosmic horizons. While the derivation itself requires advanced concepts of general relativity, some of the results are simple enough to be useful in teaching about cosmic horizons to undergraduate or even high-school students.
\end{abstract}
%\keywords{cosmological horizon, cosmic expansions, cosmology education}
%\pacs{04.20.Cv,98.80.Es,98.80.Jk}
%\submitto{\EJP}
\maketitle % title page is now complete

\section{Introduction}
\label{Sec:Introduction}
Our best current overall description of our expanding cosmos is based on the Friedmann-Lema\^{\i}tre-Robertson-Walker (FLRW) spacetimes of general relativity. In the fundamental causal structure of those models --- given that no influence can travel faster than light, which regions can influence which other regions? --- there are two different definitions of horizons, of ``causality boundaries,'' that are of interest: A {\em particle horizon} defines the boundary between spacetime regions that could have influenced the past or present of our own galaxy on the one hand, and those more distant past regions where no causal contact could have been possible on the other.  A {\em cosmic event horizon} denotes the boundary at the present (cosmic) time separating those regions whose light can reach us at some time in the future from more distant regions whose light can never reach us, no matter how long we wait.

Cosmological horizons are a recurring topic in general relativity education. Almost all of those who have participated in the debate make reference to the following intuitive picture, which comes from classical physics:  What happens when two objects move along the same line, the second object chasing the first? 

If the second object's speed is greater, it will eventually catch up with the first. If the first object's speed is greater, the second object will never catch up. It is tempting to see this as an analogy for the nature of cosmic event horizons; in the following, I will call this analogy the ``catch-up picture,'' for short. 

Consider our own galaxy and a distant galaxy, with some relative speed $v$, and consider light sent from that distant galaxy towards us at the speed of light $c$. If $v>c$, then from the point of view of the distant galaxy, its light signal will never reach us. In an expanding universe, a distant galaxy's recession speed $v_{rec}$ and its distance from us $d$ are linked via the Hubble constant $H_0$ in what is now called the Hubble-Lema\^{\i}tre law,
\be
v_{rec} = H_0\cdot d,
\ee 
the famous relation that led astronomers to realise in the late 1920s that our universe is expanding. The catch-up picture suggests that the cosmic event horizon should be located at that distance where the speed for local galaxies relative to our own reaches, or exceeds, $v=c$.

In pedagogy, simple, intuitive pictures can be a great boon. Except when an intuitive picture is fundamentally wrong, which is when it becomes a stumbling block for students seeking deeper understanding. Unfortunately, that is the case for the catch-up picture of cosmic horizons, if one takes relative speed to refer to the so-called recession speed, that is, to the speed that comes out of the Hubble-Lema\^{\i}tre law. In consequence, almost all pedagogical discussions of cosmological horizons spend significant time on pointing out the inadequacy of the catch-up picture \citep{Murdoch1977,Harrison1991,Stuckey1992,Ellis1993,DavisLineweaver2004,Davis2005,Neat2019}.

There is, however, an alternative perspective which I have not found represented in any of the published contributions to the debate on teaching about cosmological horizons. It is true that in the wide-spread {\em expanding space interpretation,} which distinguishes between ordinary motion through space and distance changes in an expanding universe \citep{DavisLineweaver2004,LineweaverDavis2005,GronElgaroy2007}, there is little choice but to understand relative speed in the catch-up picture as galaxy recession speed. But the situation is different in another, less well known interpretation of cosmic expansion: the {\em relativistic motion interpretation}, which interprets the distance changes associated with cosmic expansion as a pattern of relativistic galaxy motion, and the cosmological redshift as a general-relativistic Doppler shift \citep{Narlikar1994,Liebscher2007,BunnHogg2009,Kaya2011}.
The aim of the present article is to explore cosmic horizons from that alternative perspective. As we shall see, when it comes to cosmic horizons, the relativistic motion interpretation allows us to retain the main idea of the catch-up picture of a cosmic event horizon. The important difference is that the speeds in question are not the recession speeds from the Hubble-Lema\^{\i}tre law, but the radial relativistic velocities of galaxies --- a type of motion that is defined by using spacetime geometry to compare the four-velocities of objects, using the technique of parallel transport.

After having recapitulated the basic concepts needed to understand the FLRW models in section \ref{Sec:FLRWBasics}, I introduce relative motion in general relativity in section \ref{RelativeMotionRelativity}, and apply the concept to cosmology in section \ref{DopplerInterpretation}. Since the relativistic motion interpretation is likely to be unfamiliar to most readers, I have taken care to make the presentation self-contained. After laying out the basic properties of cosmic horizons in section \ref{Sec:CosmicHorizons}, the interpretation of cosmic horizons using a modified version of the catch-up picture for relative radial velocities is presented in section \ref{InterpretingCosmicHorizons}. While those sections make use of the full formalism of general relativity, a number of their results have a bearing on how to teach cosmology at an undergraduate or even high school level; I discuss those aspects in section \ref{Sec:Discussion}.

\section{Basics of FLRW cosmology}
\label{Sec:FLRWBasics}
To fix notation and to lay the groundwork for the calculations in section \ref{DopplerInterpretation}, this section introduces the basic elements of Friedmann-Lema\^{\i}tre-Robertson-Walker (FLRW) spacetimes. At the same time, I will introduce the basic ideas of the expanding space interpretation. The metric for FLRW spacetimes is
\be
\!\!\!\!\Dd s^2 = -c^2\Dd t^2 + a^2(t)\left[
\frac{\Dd r^2}{1-Kr^2} + r^2(\Dd\theta^2+\sin^2\theta\Dd\phi^2)
\right].
\label{FLRWMetric}
\ee
Explicit calculation shows that each set of constant values for the ``co-moving coordinates'' $r,\theta,\phi$ corresponds to a time-like geodesic of the spacetime, that is, to objects in free fall. The set of all geodesics defined in this way defines the ``Hubble flow.'' For times later than the first hundreds of millions of years of cosmic history, the Hubble flow can be pictured as a family of idealized galaxies whose mutual distances change only due to cosmic expansion.

As can be read off from (\ref{FLRWMetric}), the {\em cosmic time coordinate} $t$ is defined so that, on a Hubble-flow world-line, coordinate time intervals correspond to proper time intervals, that is, time intervals as measured on a clock carried along by the Hubble-flow galaxy in question.

FLRW universes are homogeneous and isotropic, with $K$ parametrising the three possible options for spatial curvature as constant positive for $K>0$, flat for $K=0$, and constant negative for $K<0$. The notion of simultaneity that defines cosmic time is chosen so as to respect that homogeneity: Equal values of $t$ correspond to the same local density everywhere in a FLRW universe. 

The function $a(t)$ is the cosmic scale factor, whose functional form is dependent, via the Einstein equations, on the amounts and the equations of state of the matter contained in the model. Under suitable conditions, there can be times $t$ where $\dot{a}(t)<0,$ corresponding to a collapsing universe. At the present cosmic time, the FLRW model which provides the best fit to the properties of our own universe has $\dot{a}(t)>0,$ corresponding to cosmic expansion.

Since FLRW universes are homogeneous, we are free to choose the origin of the spatial coordinate system. The common choice puts our own galaxy at $r=0$. If we then consider a galaxy in the Hubble flow whose world-line is at constant $r=r_d$, we can calculate that galaxy's proper distance from us at constant cosmic time $t$, that is, integrate up the line element $\Dd s$ given by (\ref{FLRWMetric}) along coordinate intervals $\Dd r$ from $r=0$ to $r=r_d$. It follows immediately from (\ref{FLRWMetric}) that the result is proportional to $a(t)$: Distances $d_{ij}$ between any two Hubble-flow galaxies $i,j$ in a FLRW universe change over time as
\be
d_{ij}(t) = \frac{a(t)}{a(t_0)} d_{ij}(t_0)
\label{ProperDistance}
\ee
for any arbitrary reference time $t_0$. Differentiating $d_{ij}(t)$ with respect to the cosmic time, we obtain the Hubble-Lema\^{\i}tre relation
\be
v_{ij}(t) \equiv \frac{\Dd d_{ij}}{\Dd t}(t) = \frac{1}{a(t)}\frac{\Dd a}{\Dd t}(t)\cdot  d_{ij}(t)\equiv
H(t)\cdot  d_{ij}(t),
\label{HubbleLemaitre}
\ee
where the rightmost equation introduces the Hubble parameter $H(t)$. This defines the {\em recession speed} $v_{ij}$ of the galaxy $i$ from the galaxy $j$, the rate at which the distance between those two galaxies increases over cosmic time. Applied to a single galaxy, the term commonly refers to that galaxy's recession speed from our own.

For sufficiently large distances, the recession speed of a distant galaxy defined by (\ref{HubbleLemaitre}) will become faster than the speed of light. For given $H(t)$, the distance value at which the recession speed reaches the speed of light defines the radius of the so-called {\em Hubble sphere} at cosmic time $t$. The superluminal recession speeds beyond that distance are bound to alarm students who remember how Einstein's special theory of relativity tells us that nothing can move faster than the speed of light. 

This is where the {\em expanding space interpretation} comes into play. At the heart of that interpretation is an explicit distinction between motion through space on the one hand and cosmic expansion on the other. The fact that Hubble-flow galaxies are at rest in co-moving coordinates is extended to the more general statement that Hubble-flow galaxies are at rest in the expanding universe. In consequence, the increasing distances between them are attributed not to motion through space, but instead to changes in the properties of space between those galaxies, to an ``expansion of space.'' 

Worries about superluminal recession speeds are dismissed by stating that special relativity's speed limit applies only to the motion of objects through space, while in an expanding universe, it is space itself that expands; distance changes of this kind, it is stated, are not subject to special relativity's restriction.

Light emitted at a wavelength by a distant Hubble-flow galaxy and observed as it arrives in our own galaxy is redshifted. Specifically, let 
\be
z=\frac{\lambda_r-\lambda_e}{\lambda_e}
\ee
be the redshift, where $\lambda_e$ is the wavelength at which the light has been emitted as measured by a local observer in the distant galaxy, and $\lambda_r$ the wavelength we measure for the arrival of the light. For small recession speeds, $v_j\ll c,$ the cosmological redshift for light arriving at time $t=t_0$ is given by the classical Doppler shift
\be
z = \frac{v_j(t_0)}{c}
\ee
for the galaxy's present-day recession speed $v_j$. But this is only an approximation; the exact formula can be calculated to link the redshift directly to the scale factor values at the time $t_e$ the light was emitted and the time $t_0$ it was received, as
\be
1+z = \frac{a(t_0)}{a(t_e)}.
\label{CosmologicalRedshift}
\ee
In the expanding space interpretation, this direct connection, with wavelengths increasing in exactly the same way as inter-galaxy distances in the Hubble flow, is taken to indicate that the cosmological redshift is caused directly by the expansion of space \citep{Seeds2007,DavisLineweaver2004}. The interpretation of the cosmological redshift as a Doppler effect, it is argued, needs to be replaced by the ``more accurate view [that light] waves are stretched by the stretching of space'' \citep{Fraknoi2004}.

\section{Relative motion in relativity}
\label{RelativeMotionRelativity}

We have seen how the expanding space interpretation deals with superluminal recession speeds by stating that they do not correspond to motion through space. The road to the alternative relativistic motion interpretation of cosmic expansion in FLRW models with metric (\ref{FLRWMetric}) begins with the closely related question: What is motion (through space) in a general-relativistic universe in the first place? 

The recession speeds (\ref{HubbleLemaitre}) are obtained by measuring a distance along a hypersurface of constant coordinate time, then taking the derivative with respect to that coordinate time. This reliance on coordinates, even for a coordinate system that is particularly well suited to the symmetries of the underlying spacetime, is alien to relativity, where we learn as early as special relativity that the proper way to talk about relative motion involves the four-velocities associated with the world-lines of objects. 

In search of a more appropriately relativistic definition, let us recapitulate some basic facts about the notion of relative velocity in special and in general relativity, following in the footsteps of \cite{Synge1966}, \cite{Narlikar1994}, and \cite{BunnHogg2009}.

In special relativity, in order to determine the relative velocity for two objects 1 and 2, you will need to specify which two events ${\cal E}_1$ and ${\cal E}_2,$ one on each object's world-line, you want to include in your comparison. Find the momentarily co-moving inertial frame $\cal I$ for object 2 at ${\cal E}_2.$ Then, determine the four-velocity of object 1 at event ${\cal E}_1$ in the inertial frame $\cal I$. In the usual \mbox{(pseudo-)}Cartesian coordinates and with the Minkowski metric $\eta=\mbox{diag}(-1,+1,+1,+1),$ the four-velocity will have the form
\be
\fvec{u}=\gamma(v)\left(
\begin{array}{c}
c\\
v_x\\
v_y\\
v_z
\end{array}
\right),
\label{GeneralFourVelocity}
\ee 
where $v_x, v_y$ and $v_z$ define the three-velocity, which corresponds to a speed $v=\sqrt{v_x^2+v_y^2+v_z^2}$, where $\gamma(v)\equiv [1-(v/c)^2]^{-1/2},$ and where we introduce the convention of denoting four-vectors by underlined letters. That the time-time component of the Minkowski metric is $-1$ instead of $-c^2$ is a clash between the usual conventions for special and for general relativity; where $c$ is already contained in the metric itself, as in (\ref{FLRWMetric}), the time component of the four-velocity would just be $\gamma(v)$. 

\begin{figure}[htbp]
\begin{center}
% Preprint version
%\includegraphics[width=0.45\textwidth]{figure-four-velocities.pdf}
\includegraphics[width=0.75\linewidth]{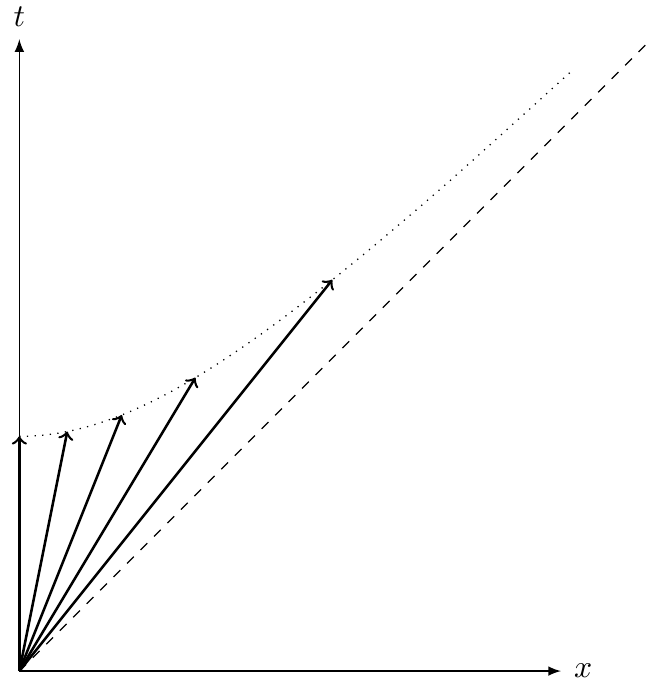}
\caption{Four-velocity vectors for $v_x/c=0, 0.2, 0.4, 0.6, 0.8$ in an $xt$ spacetime diagram with units chosen so that $c=1$. The hyperbola $(u^0)^2-(u^1)^2=c^2$ is shown as a dotted line, the light cone as a dashed line }
\label{Fig:FourVelocities}
\end{center}
\end{figure}

In a spacetime diagram, four-velocity vectors look increasingly ``long'' as the three-velocity increases. For motion in one dimension, the x direction, this is shown in Fig.~\ref{Fig:FourVelocities}. In the diagram, the tips of the four-velocity vectors lie on the hyperbola $(u^0)^2-(u^1)^2=c^2$.

The three-vector $\vec{v} = (v_x,v_y,v_z)^T$ determines the velocity of the second object relative to the first. If you are only interested in the magnitude $v$, that is, in the relative speed, there is a quick coordinate-independent way of finding the answer, which follows directly from the preceding argument and from the fact that, in its own momentary rest system $\cal I$, the four-velocity of object 1 is $\fvec{w}=(c,0,0,0)^T$: Calculate
\be
\gamma(v) = -\eta(\fvec{w},\fvec{u})/c^2
\ee
and solve for $v$. For radial motion, where object 1 is moving directly away from, or directly towards object 2, we can define the object 1's radial velocity relative to object 2 as $v_R=\pm v,$ where the plus sign indicates motion away from object 2 and the minus sign motion towards object 2. 

In order to determine the wavelength shift for a light signal sent from object 1 to object 2 due to the Doppler effect, we apply this procedure to two specific events: The event ${\cal E}_1$ at which the light was emitted by object 1 and the event ${\cal E}_2$ at which the light is received by object 2. Under those circumstances, the Doppler shift $z$ related to the radial velocity $v_R$ for the two objects, evaluated for the emission and reception events, is given by the special-relativistic (longitudinal) Doppler formula for radial motion as
\be
1+z = \sqrt{\frac{1+v_R/c}{1-v_R/c}} \equiv k(v_R),
\label{SRDoppler}
\ee 
where the rightmost equation defines the {\em Bondi k factor} \citep{Bondi1963}.

In generalising this definition of relativistic relative velocity to general relativity, the key problem is that, in a curved spacetime, four-vectors are only defined locally. At each event, there is a corresponding tangent space of local four-vectors. In our special-relativistic calculation, we have taken a vector defined at one event and compared it, component by component, to a vector defined at another event. In general relativity, those two vectors are not even part of the same mathematical vector space.

The same problem is encountered when it comes to those comparisons of four-vectors, or of more general tensors, that are necessary in order to define the spacetime derivative of a vector field. Taking the derivative amounts to comparing the value of a field at one event with the value at a neighbouring event --- and again, in a general-relativistic spacetime, a vector defined at one event cannot be compared directly with a vector defined at a different event, even if the two events are infinitesimally close.

The additional structure that is needed is one that connects different tangent spaces: a linear connection. In general relativity, this is the Levi-Civita connection. Given two events that are linked by a curve, the connection tells us how to transport vectors from the first event along the curve to the second event. Since this kind of transport is the closest we can get to not changing the direction of the vectors while moving them from one event to another for comparison purposes, it is commonly called {\em parallel transport}. Comparison of infinitesimally close vectors, or tensors, by parallel transport, is at the heart of the so-called covariant derivative, general relativity's tool of choice for quantifying physical change.

The Levi-Civita connection is a metric connection, meaning that it respects scalar products: if $\fvec{u},\fvec{w}$ are four-vectors at some event ${\cal E}_1$, and $\fvec{u}',\fvec{w}'$ are the results of parallel-transporting $\fvec{u}$ and $\fvec{w}$ along the same curve to another event ${\cal E}_2$, then we have
\be
\left. g(\fvec{u},\fvec{w})\right|_{{\cal E}_1} =  \left. g(\fvec{u}',\fvec{w}')\right|_{{\cal E}_2}.
\ee
The straightest-possible curves are those whose tangent vectors experience no physical change along the curve --- in other words: the tangent vector at each point of that curve is obtained by parallel-transporting the initial tangent vector to that point, along the curve itself. Those straightest-possible curves are the {\em geodesics} associated with the metric $g$. In general relativity, the world-lines of test particles in free fall are geodesics, as are the world-lines of photons. Geodesics can be found by solving the geodesic equation that encodes the parallel-transport-along-itself of the tangent vector; if the curve is given by $x^{\mu}(\lambda)$, the equation is
\be
\frac{\Dd^2 x^{\mu}}{\Dd\lambda^2}
+\Gamma^{\mu}_{\rho\nu} \frac{\Dd x^{\rho}}{\Dd\lambda}\frac{\Dd x^{\nu}}{\Dd\lambda} = 0,
\label{GeodesicEquation}
\ee
where $\lambda$ is called an {\em affine parameter} and where the {\em Christoffel symbols (of the second kind)} $\Gamma^{\mu}_{\rho\nu}$ that make up the connection are defined as
\be
\Gamma^{\mu}_{\rho\nu} = \frac12 g^{\mu\sigma}\left[
\frac{\pa g_{\nu\sigma}}{\pa x^{\rho}}
+\frac{\pa g_{\rho\sigma}}{\pa x^{\nu}}
-\frac{\pa g_{\rho\nu}}{\pa x^{\sigma}}
\right].
\label{Christoffel2}
\ee
Once we have parallel-transported the four-velocity of object 1 to ${\cal E}_2$, the equivalence principle tells us that we can find a local inertial frame, co-moving with object 2. In that frame, the parallel-transported four-velocity has the form (\ref{GeneralFourVelocity}), from which we can directly read off the components of the relative velocity of object 1 at ${\cal E}_1$ relative to object 2 at ${\cal E}_2$.

In general, the result of parallel transport will depend on the curve along which the vector in question has been transported. This is a direct consequence of the curvature of spacetime, and in fact the path-dependence of parallel transport is at the core of the definition of spacetime curvature via the so-called Riemann curvature tensor.  The path-dependence makes the comparison of four-velocities, and thus the definition of a relative velocity, more ambiguous in general relativity. 

There are several possible takes on this. One could argue that, given this ambiguity, the concept of relative velocity simply cannot be generalised properly to general relativity \citep{Kaya2011}. Alternatively, we can accept that in general relativity, in order to define relative velocity, we need to specify more than in special relativity: not only two world-lines, and one event on each, but also the spacetime curve linking the two events. 

That said, there is a natural way of reducing the ambiguity by only considering those transport-curves that are geodesics. This is analogous to the general-relativistic generalisation of the spatial distance between two events whose separation is space-like. In that case, as well, we must realise that we cannot just specify ``the spatial distance'', since spatial distance has become path dependent. But this is commonly not seen as a reason to get rid of the concept of spatial distance altogether. Instead, we can find the geodetic segment linking the two events, and use it to assign a coordinate-independent distance value. In a similar vein, we define relative velocity by parallel transport along geodesic segments.

Taking all this together, we have arrived at the following definition: Given the world-line $w_1$ of object 1, and an event ${\cal E}_1$ on that world-line, and the world-line $w_2$ of object 2, and an event ${\cal E}_2$ on that second world-line, we define the velocity of object 2 at ${\cal E}_2$ relative to object 1 at ${\cal E}_1$ by transporting the four-velocity vector of $w_1$ at ${\cal E}_1$ along the geodesic segment linking ${\cal E}_1$ and ${\cal E}_2$, and then evaluating the resulting four-vector at ${\cal E}_2$ in the momentarily co-moving local inertial frame of $w_2$ according to (\ref{GeneralFourVelocity}). 

The modified definition has the advantage of being unique at least in a neighbourhood of each event --- each event $\cal E$ has a neighbourhood of other events that are linked to $\cal E$ by a unique segment of a geodesic \citep{Bolos2007}. On the other hand, the resulting generalised relative velocities have some additional properties that go beyond special relativity. Notably, once this definition is chosen, spectral shifts in static gravitational fields can always be interpreted as Doppler shifts due to the relative velocity of the observers involved, even though the observers in question are both at rest in a coordinate system adapted to the staticity of the spacetime. Erwin Schr\"odinger appears to have been the first to point out that ambiguity between gravitational frequency shifts and Doppler shifts in his 1954 lectures on expanding universes \citep{Schroedinger1956}. John Lighton Synge went so far as to argue that, in consequence, what is commonly called the gravitational redshift should not be classed as an effect of gravity at all, since the Riemann tensor does not appear in the relevant formulae --- the gravitational redshift is not associated with curvature, but with that portion of classical gravity that is only present for a suitable choice of coordinates \citep{Synge1966}.

For completeness, let us note in passing that there are also definitions of relative velocity using parallel transport that do {\em not} generalise the two-event definition of special relativity \citep{Bolos2007,Chodorowski2011}, but since the aim here is to generalise the special-relativistic version of the concept, I will not consider those alternative definitions in the following. 

\section{The relativistic motion interpretation of cosmic expansion}
\label{DopplerInterpretation}

After these preparations, we can come back to the question of how to properly define the relative velocities of galaxies in an expanding FLRW universe. While we have seen in the previous section that velocity comparisons based on parallel transport have an inherent ambiguity in curved spacetime, it is worth noting that, when we are talking about the cosmological redshift, there necessarily {\em is} a preferred curve by construction, a geodesic no less, linking the two events in question: the geodesic along which the light signals travels from the emission event in the distant galaxy to the reception event in our own. Whenever we are talking about galaxies so far away that light from a certain event on their world-line cannot reach us, on the other hand, there is no preferred (space-like) geodesic from that specific event, intersecting our world-line; we will leave this latter situation out of our considerations, and concentrate on emission events that have a corresponding reception event on our own world-line.

\begin{figure}[htbp]
\begin{center}
% Material/paralleltransport-abbildung.tex
%Preprint version
%\includegraphics[width=0.45\textwidth]{figure-parallel-transport.pdf}
\includegraphics[width=0.6\linewidth]{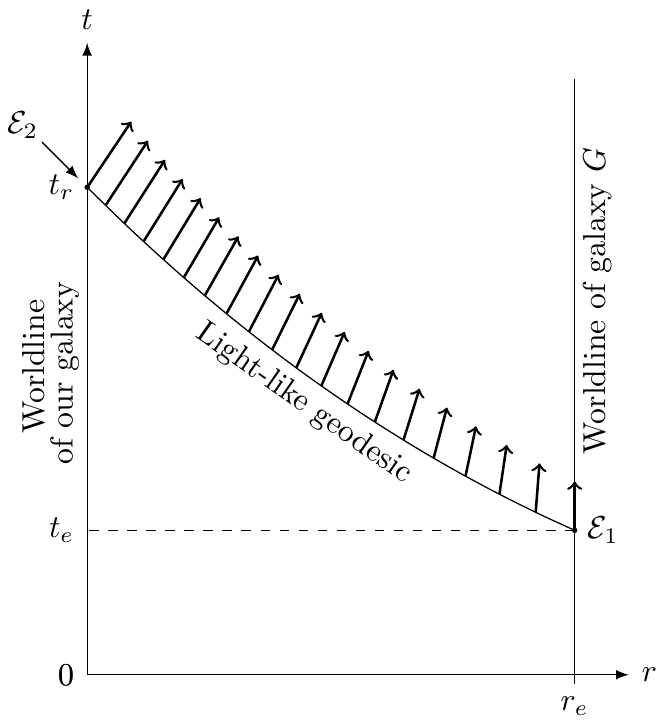}
\caption{The basic setup for transporting the four-velocity of a distant galaxy $G$ along the light-like geodesic associated with light reaching us from that galaxy. Co-moving coordinates corresponding to physical distances at the reception time (present time), units so that $c=9$, in an Einstein-de Sitter universe with $a(t)\sim t^{2/3}$. Also shown are several parallel-transported versions of the four-velocity of galaxy $G$, 
represented by their components in the local orthonormal frame at rest in the Hubble flow
}
\label{Fig:ParallelTransportSetup}
\end{center}
\end{figure}

What, then, is the relative velocity of the world-line of the distant Hubble-flow galaxy $G$ and our own galaxy, evaluated at the emission event ${\cal E}_1$ on the world-line at $G$, and the reception event ${\cal E}_2$ for the same light on our own galaxy's world-line, evaluated by parallel transport along the light's geodesic? The setup, and some intermediate states of the parallel-transported four-velocity of the galaxy $G$, are shown in Fig.~\ref{Fig:ParallelTransportSetup}, where four-vectors are plotted using their components within a local orthonormal system that is at rest in the Hubble flow. As expected from Fig.~\ref{Fig:FourVelocities}, while the four-velocity is transported towards ${\cal E}_1$, the radial velocity $v_R$ as measured by a local observer becomes larger, and so in the spacetime diagram, the corresponding vector becomes longer and increasingly more inclined.

The following argument combines steps outlined by \cite{Narlikar1994} with the more general definition given by \cite{Bolos2007}. Let the coordinates of ${\cal E}_1$ be $r=r_e$ and $t=t_e$. The reception event ${\cal E}_2$ happens at the location of our own galaxy, at $r=0$, and at the time $t_0$ that, in cosmology, typically refers to the present.

Let us denote the light-like geodesic linking the two events by $c(\lambda)$, with $\lambda$ a suitable affine parameter, and let us call its
tangent vector $\fvec{L}(\lambda)$, with components
\be
L^{\mu}(\lambda) = \frac{\Dd c{\,}^{\mu}}{\Dd\lambda}(\lambda).
\ee
For radial motion, the only non-zero components are the time component $L^0(\lambda)$ and the radial component $L^1(\lambda)$. From the geodesic equation (\ref{GeodesicEquation}), and plugging in the connection coefficients for the FLRW metric (\ref{FLRWMetric}), we have 
\be
\frac{\Dd L^0}{\Dd\lambda} + \frac{a\cdot\dot{a}}{c^2(1-Kr^2)}\left(L^1(\lambda)\right)^2=0,
\label{TimeGeodesicFLRW}
\ee
(cf. appendix \ref{App:Christoffel}), where the dot denotes differentiation with respect to $t$. For a light-like geodesic, we have the additional condition that $\Dd s^2=g(\fvec{L},\fvec{L})=0$, which for the particular case of a radial geodesic ($\Dd\phi=0$ and $\Dd\theta=0$) for inward-travelling light ($\Dd r<0$) means
\be
\frac{1}{\sqrt{1-Kr(\lambda)^2}}L^1(\lambda) = -\frac{c}{a[t(\lambda)]}L^0(\lambda).
\label{DrDtRelationGeodesic}
\ee
We can use this result to rewrite (\ref{TimeGeodesicFLRW}) as a differential equation for $L^0$ only, namely
\be
\frac{\Dd L^0}{\Dd\lambda} +\frac{\dot{a}}{a}\left(L^0(\lambda)\right)^2 = 0.
\label{Eq:LambdaZeroEquation}
\ee
But $\dot{a}\cdot L^0$ is the derivative of $a$ with respect to $\lambda,$ since 
\be
 \frac{\Dd a}{\Dd\lambda} = \frac{\Dd a}{\Dd t}\cdot \frac{\Dd t}{\Dd\lambda} = \dot{a} L^0, 
\ee
where the derivative of $t$ with respect to $\lambda$ is taken along the geodesic, and thus corresponds to the time component of the geodesic's tangent vector. With this in mind, equation (\ref{Eq:LambdaZeroEquation}) is readily integrated with respect to cosmic time $t$ to yield
\be
a\cdot L^{0} = C,
\label{AffineFactorEquation}
\ee
with $C$ some integration constant. The affine parameter is only fixed up to an affine transformation $\lambda\to\lambda'=f\lambda+g$ for constants $f,g$, so we can choose to have our geodetic segment start with $\lambda=0$ at ${\cal E}_2$ and end with $\lambda=1$ at ${\cal E}_1$. This fixes the integration constant as 
\be
\int\limits_{t_e}^{t_0}a(t)\:\Dd t = \int\limits_0^1 C\:\Dd\lambda = C,
\ee
where we have applied separation of variables, with $t$ on the left and $\lambda$ on the right, to (\ref{AffineFactorEquation}). In conclusion, we can write down the parameter-dependent tangent vector of our geodetic segment as
\be
\fvec{L}(\lambda) = \frac{C}{a(\lambda)^2}\left(
\begin{array}{c}
a(\lambda) \\[0.3em]
-c\sqrt{1-Kr^2}\\[0.1em]
0\\
0
\end{array}
\right),
\label{TangentVectorLambda}
% PN8 15
\ee
whence the components $L(\lambda)^{\mu}$ can be read off directly.

Next, define the four-vector-valued function $\fvec{u}(\lambda)$, where $\fvec{u}(0)$ is the four-velocity of the galaxy $G$ at the emission event ${\cal E}_1$, and where $\fvec{u}(\lambda)$ the result of parallel-transporting $\fvec{u}(0)$ along our light-like geodesic segment to the event corresponding to the parameter value $\lambda$. In particular, $\fvec{u}(1)$ is the result of parallel-transporting $\fvec{u}(0)$ to the reception event of the light, in our own galaxy. For reasons of symmetry, $\fvec{u}(\lambda)$'s only non-zero components are in the time and radial directions. We can readily see that $\fvec{u}(0) = (1,0,0,0)^T$, since $G$ is in the Hubble flow and thus at rest in our chosen coordinate system, and since our chosen time coordinate is the time coordinate of the local co-moving inertial system at ${\cal E}_1$.

The fact that parallel transport preserves scalar products saves us from having to solve another set of differential equations. Instead, $g(\fvec{L}(0),\fvec{u}(0)) \stackrel{!}{=}g(\fvec{L}(1),\fvec{u}(1))$ and $g(\fvec{u}(0),\fvec{u}(0))=g(\fvec{u}(1),\fvec{u}(1))=-c^2$ allow us to solve for the non-zero components of $\fvec{u}(1)$ algebraically, as
\be
u^0(1) = \frac12\left[
\frac{a(t_e)}{a(t_0)}+\frac{a(t_0)}{a(t_e)}
\right]
\label{u0Determined}
\ee
and 
\be
u^1(1) = \frac{c}{2a(t_0)}\left[
\frac{a(t_0)}{a(t_e)}-\frac{a(t_e)}{a(t_0)}
\right].
\ee
The cosmological coordinate system at ${\cal E}_2$ is co-moving with our own galaxy, and the vectors of the time and radial directions, $\fvec{e}'{}_t=(1,0,0,0)^T$ and $\fvec{e}'{}_r=(0,1,0,0)^T$, are orthogonal to each other. But these vectors are not yet ortho{\em normal}, since $g(\fvec{e}'{}_t,\fvec{e}'{}_t)=-c^2$ and $g(\fvec{e}'{}_r,\fvec{e}'{}_r)= a(t_0)^2$. In order to express $\fvec{u}(1)$ in terms of the unit vectors of a local inertial frame, we need to know its components in terms of re-scaled basis vectors $\fvec{e}{\,}_t\equiv \fvec{e}'{}_t/c$ and $\fvec{e}{\,}_r\equiv \fvec{e}'{}_r/a(t_0)$. In terms of those orthonormal basis vectors, we have
$\fvec{u}(1) = \bar{u}^0 \cdot \fvec{e}{\,}_t + \bar{u}^1\cdot \fvec{e}{\,}_r$, and from comparing these components with the 
generic form (\ref{GeneralFourVelocity}), we can read off that the radial relative velocity is
\be
v_R = c\frac{\bar{u}^1}{\bar{u}^0} = c\left(\frac{a(t_0)^2-a(t_e)^2}{a(t_0)^2+a(t_e)^2}\right). 
\label{RadialVelocityCosmo}
\ee
Note that with this definition, $v_R$ can be arbitrarily close to $c$, but never larger: $v_R< c$, a fact that can not only be derived from the explicit form for $v_R$ but, more abstractly, from the fact that parallel-transport with a metric connection preserves the property of a four-vector to be time-like. The corresponding Doppler shift, which can be obtained by applying the special-relativistic Doppler formula
(\ref{SRDoppler}) in the local co-moving inertial system, is
\be
1+z = k(v_R) = \frac{a(t_0)}{a(t_e)}.
\label{CosmologicalDoppler}
\ee
Thus, the Doppler shift derived from the parallel-transported radial relative velocity 
reproduces exactly the general formula (\ref{CosmologicalRedshift}) for the cosmological redshift in terms of cosmic scale factor values. 

In this way, the relativistic motion interpretation provides a consistent picture of cosmology: cosmic expansion means that galaxies in the Hubble flow are moving away from us; their radial velocities remain subluminal, and the cosmological redshift is explained by the Doppler effect corresponding to their radial motion --- obtained by plugging the relative radial velocity we have determined via parallel transport into the usual special-relativistic Doppler formula. The Doppler interpretation has the added advantage that there is nothing surprising about photons being observed at the reception event with larger wavelength, and thus with lower energy, than the one at which they were emitted. In the expanding space interpretation, without reference to a Doppler effect, the apparent energy loss of redshifted photons is less straightforward to explain, and requires recourse to unintuitive statements about global energy not being conserved in general relativity \citep{Davis2010}. 

\begin{figure}[htbp]
\begin{center}
% Material/paralleltransport-abbildung.tex
% Preprint version
%\includegraphics[width=0.45\textwidth]{figure-parallel-transport-peculiar.pdf}
\includegraphics[width=0.64\linewidth]{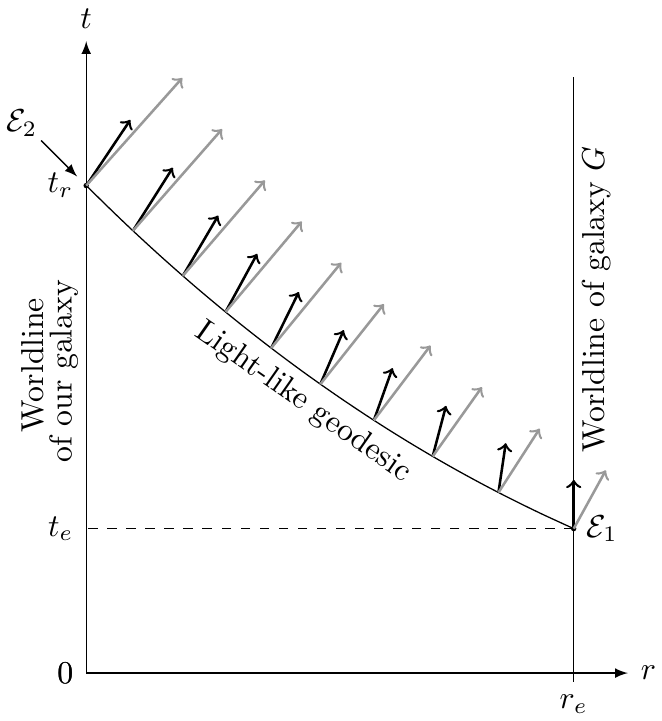}
\caption{Transporting the four-velocity of a distant galaxy $G$ and an entity $E$ with non-zero peculiar velocity along the light-like geodesic associated with light reaching us from that galaxy. Co-moving coordinates corresponding to physical distances at the reception time (present time), units chosen so that $c=9$, in an Einstein-de Sitter universe with $a(t)\sim t^{2/3}$. Four-vectors are again represented by their components in the appropriate local co-moving orthonormal frame. The peculiar velocity of $E$ relative to $G$ is $w_P=0.55\:c$
}
\label{Fig:ParallelTransportSetup2}
\end{center}
\end{figure}

What about an entity (object or light signal) $E$ that is present at the event ${\cal E}_1$, but which has non-zero (peculiar) radial velocity relative to the Hubble-flow galaxy $G$? This case has been treated in an article by \cite{Emtsova2020}; the following is a derivation within the framework described above. The setup is shown in Fig.~\ref{Fig:ParallelTransportSetup2}. This time, two four-vectors are being parallel-transported, namely the four-velocities of galaxy $G$ (black arrows) and that of the entity $E$ (grey arrows).

Building on the previous definitions in this section, define the four-vector-valued function $\fvec{w}(\lambda)$ where $\fvec{w}(0)$ is the four-velocity of $E$ at the event ${\cal E}_1$ and $\fvec{w}(\lambda)$ is that initial four-velocity, parallel-transported along the light-like geodesic joining ${\cal E}_1$ and ${\cal E}_2$ to the event corresponding to the parameter value $\lambda$. In order to write down $\fvec{w}(0)$ in terms of the relative radial velocity $\fvec{w}_P$ of $E$ and $G$ at ${\cal E}_1$ (which is the peculiar velocity, in cosmological terms), we introduce a local inertial coordinate system at ${\cal E}_1$ that is at rest relative to $G$. We only need the unit vectors for the time and radial directions, which, following the same steps by which we found the local inertial coordinate system at ${\cal E}_2$, are $\fvec{\hat{e}}{\,}_t=(1/c,0,0,0)^T$ and $\fvec{\hat{e}}{\,}_r=(0,\sqrt{1-Kr_e^2}/a(t_e),0,0)^T$. From the generic form (\ref{GeneralFourVelocity}) of the four-velocity in special relativity, it follows that the four-velocity $\fvec{w}(0)$ in that local inertial coordinate system must be related to the relative radial speed $w_P$ as
\be
\fvec{w}(0) = \gamma(w_P)\left[
c\:\fvec{\hat{e}}{\,}_t + w_P\: \fvec{\hat{e}}{\,}_r
\right] =
\left(
\begin{array}{c}
\gamma(w_P)\\
\gamma(w_P)\:w_P\frac{\sqrt{1-Kr_e^2}}{a(t_e)}\\
0\\
0
\end{array}
\right).
\label{w0Form}
\ee
As in the previous derivation, invariance of scalar products under parallel transport, namely $g(\fvec{L}(0),\fvec{w}(0)) \stackrel{!}{=}g(\fvec{L}(1),\fvec{w}(1))$ and $g(\fvec{w}(0),\fvec{w}(0))=g(\fvec{w}(1),\fvec{w}(1))=-c^2,$ allows us to solve for $\fvec{w}(1)$ algebraically. Notably, all the factors containing the spatial curvature $K$ cancel; the result is
\bea
w^0(1)&=&\frac12\left[ k(v_R)\cdot k(w_P) + \frac{1}{k(v_R)\cdot k(w_P)}  \right]\\[1em]
w^1(1)&=&\frac{c}{2a(t_0)}\left[k(v_R)\cdot k(w_P) - \frac{1}{k(v_R)\cdot k(w_P)}\right],
\eea
where the use of Bondi k-factors makes for the most compact notation for the resulting components. As we did for the galaxy $G$, we can express this four-velocity in the local inertial system co-moving with our own galaxy, in order to derive the relative radial velocity $v_{PR}$ of $E$ and our own galaxy. The result is 
\be
v_{PR} = \frac{a(t_0)\cdot w^1(1)}{w^0(1)} = \frac{v_R+v_P}{1+v_R\cdot v_P/c^2}.
\label{VelocityAddition}
% PN5 106
\ee
This is the special-relativistic addition formula for parallel velocities, and thus provides an important consistency check for the relativistic motion interpretation of the cosmological redshift. Not only does parallel transport of velocities provide for a Doppler interpretation for the cosmological redshift of a Hubble-flow galaxy, but also the resulting relative radial velocity combines with peculiar radial velocities in the proper way. By way of comparison, note that the recession speed defined in the first part of (\ref{HubbleLemaitre}) has the unintuitive property that even a galaxy whose peculiar velocity exactly compensates for its recession speed still has non-vanishing redshift \citep{Kiang2001}. In contrast, it follows directly from the relativistic velocity-addition formula that a galaxy whose peculiar velocity exactly balances its radial relative velocity has cosmological redshift zero. In both cases, the relative radial velocity conforms more closely with classical preconceptions of the properties of relative motion than the recession speed.

\section{Cosmic horizons}
\label{Sec:CosmicHorizons}

The existence (or not) of horizons in FLRW spacetimes can be deduced from the propagation of light.\footnote{The derivations in question can be found in most text books covering relativistic cosmology, e.g. section 1.13 in \cite{Weinberg2008}  or section 17.3 in \cite{Rindler2001}. A detailed treatment can be found in \cite{Ellis2015}.
} The basic equation follows directly from the condition $\Dd s^2=0$ for light-like geodesics, which we had already exploited in (\ref{DrDtRelationGeodesic}). With the definitions of the previous two sections, a light signal emitted by a Hubble-flow galaxy with radius value $r_1$ at cosmic time $t_1$ and travelling radially to a second Hubble-flow galaxy at $r_2$, arriving at time $t_2$, satisfies
\be
 \int\limits^{t_2}_{t_1}\frac{c\:\Dd t}{a(t)} =\pm \int\limits^{r_2}_{r_1}\frac{\Dd r}{\sqrt{1-Kr^2}}.\label{LightSignalIntegral1}
\ee
The absolute value of the expression on the right is the definition of the {\em comoving distance} between the two galaxies,
\be
 d_{cm}(r_1,r_2)\equiv \left|  \int\limits^{r_2}_{r_1}\frac{\Dd r}{\sqrt{1-Kr^2}}\right|.
 \label{ComovingDefNew}
\ee 
Multiply the comoving distance with the current value $a(t_0)$ of the scale factor, and you will obtain the {\em proper distance} (at the present time), which can be obtained by using the FLRW metric to integrate up the length of the line joining our own galaxy and $G$ at constant cosmic time $t_0$. The plus sign is valid for $r_2>r_1$ (radially outward motion), the minus sign for $r_2<r_1$ (inward motion).

The {\em particle horizon} is defined as the boundary of influence from the past: What is the distance of the most distant cosmic location whose signals travelling at light speed can reach us at the present time? Equivalently: Which are the most distant regions in the cosmos we can observe today? At the very earliest, any light signals travelling towards us from a distant location could only have been sent out at the time of the Big Bang, $t_e=0$. Inserting this into eq.~(\ref{LightSignalIntegral1}), we obtain the current distance to the particle horizon. In the distant future, at larger values of the cosmic time coordinate $t$, our distance from the particle horizon will have become larger, as well. If you take into account that light could not travel undisturbed through the plasma of the early universe, and has only been able to stream freely since the time of recombination (according to our best current estimates, 380~000 years after the Big Bang), the analogous calculations for $t_e= $380~000 years will yield the distance to the boundary, and thus the size, of the currently observable universe.

In this article, we are primarily concerned with the second kind of cosmic horizon, commonly referred to as a {\em cosmic event horizon}. Consider all the light signals sent out at the present moment $t_0$ of cosmic time, from galaxies throughout the universe. Depending on the values of the cosmic parameters that determine the functional form of $a(t)$, there are FLRW models in which signals sent out from galaxies out to some co-moving distance $d_H$ will reach us some time in the future, while signals sent from distances beyond $d_H$ will never reach us. In those cases, the boundary sphere at $d_H$ is a cosmic event horizon.

In FLRW models, there are two basic types of cosmic event horizon. The first type is found in any universe with a finite overall life-time --- a universe that expands from a Big Bang state, reaches a maximal scale factor value at some cosmic time $t_{max}$, and collapses to a Big Crunch $a(t_{end})=0$ at some finite cosmic time $t_{end}$. In that case, the maximal distance of distant galaxies whose light, sent out at the present value $t_0$ of cosmic time, still reaches us, is given by
\be
d_H = \int\limits^{t_{end}}_{t_0}\frac{c\:\Dd t}{a(t)},
\ee
using again eq.~(\ref{LightSignalIntegral1}), with the time values chosen to indicate that this is light emitted at the present cosmic time $t_0$ which reaches us at the latest-possible cosmic time, $t_{end}$. Let us call this a {\em finite-time cosmic event horizon}.

The second type of cosmic event horizon is the one of most interest in the context of this paper. Let us call this an {\em infinite-time cosmic event horizon}.
 It can be found in certain universes that are infinitely long-lived, and that keep expanding as $t\to\infty$. In such a universe, at cosmic time $t_0$, the co-moving distance from us to the cosmic event horizon is
\be
d_H = \int\limits^{\infty}_{t_0}\frac{c\:\Dd t}{a(t)}.
\label{ITCEH}
\ee 
Whether this kind of horizon exists depends on the convergence properties of the integral on the right-hand side. If the integral diverges, then there is no finite distance value $d_H$, and thus no cosmic event horizon. If, on the other hand, the integral converges, we obtain a finite value for $d_H$, and there is indeed an event horizon, separating distant regions whence no signals can reach us from nearer regions, whose signals will reach us some time in the future.

Among the FLRW models which contain radiation, pressure-less matter, and dark energy as a non-interacting mix, those where the cosmos expands indefinitely either have a universal scale factor that, asymptotically, goes as $a(t)\sim t^q$ for some rational $q$, or, once dark energy dominates, approach a de Sitter-like universe with $a(t)\sim \exp(H'\cdot t)$. A de Sitter universe has an infinite-time cosmic event horizon at $d_H=  c/H'\cdot\exp(H'\cdot t_0)$, while the other universes have such a horizon iff $q>1$ \citep{Ellis2015}.

\section{Interpreting cosmic horizons}
\label{InterpretingCosmicHorizons}

The calculations of the previous section require familiarity with basic concepts of the cosmological models, up to and including formula (\ref{LightSignalIntegral1}) governing the propagation of light. In less advanced settings, teaching about horizons requires a more descriptive approach. If students are to understand what is going on, we will need to introduce qualitative reasoning based on simple principles of physics.

For the particle horizon and the finite-time cosmic event horizon, such reasoning is easy to come by: Light travels at a finite speed; since the moment of the Big Bang, a finite amount of cosmic time has passed; in a finite interval of time and at finite speed, light can only travel a finite distance. For light reaching us now, at this moment, that distance is the current particle horizon. For light emitted at the present cosmic time $t_0$, and arriving at our location at the Big Crunch, $t_{end}$, that distance is the finite-time cosmic event horizon.

For infinite-time cosmic event horizons, there is a natural qualitative explanation based on the intuition of classical physics, namely the catch-up section from section \ref{Sec:Introduction}: Consider the speed of our own galaxy relative to a distant light-emitting galaxy. From the perspective of an observer on that distant galaxy, it is our own galaxy that is moving away, and the light signals are chasing us. If our galaxy is as fast as those light signals (or possibly faster), the light signals will never reach us. In an expanding cosmos, relative speed of Hubble-flow galaxies grows with distance. In this interpretation, the infinite-time cosmic event horizon is that distance $d_H$ so that the radial velocity of a galaxy at that distance, relative to our own, reaches the speed of light. Light emitted by a galaxy at that distance will chase us in vain, never gaining any ground on us.

Whether or not that intuitive picture is valid depends on your definition of speed. For the recession speed  (\ref{HubbleLemaitre}) that follows from the Hubble-Lema\^{\i}tre relation, the argument is wrong. The Hubble sphere, as the sphere of galaxies with recession speeds $c$, does not coincide with the cosmic event horizon. In the framework of the relativistic motion interpretation of the cosmological redshift, on the other hand, talking about the relative radial speed we defined in section \ref{RelativeMotionRelativity} and calculated for FLRW spacetimes in section \ref{DopplerInterpretation}, the catch-up picture is a helpful short-hand description which does not lead to contradictions. 

To see this, consider any universe with an infinite-time cosmic event horizon at finite comoving distance $d_H$. The horizon distance $d_H$ is fixed by the integral on the right-hand side of eq.~(\ref{ITCEH}), which must be convergent in this case. For the integral to be convergent, we must necessarily have $\lim\limits_{t\to\infty} a(t) = \infty$. But for the light signal emitted from a galaxy at comoving distance $d_H$ at time $t_0$, and reaching our galaxy in the infinite future, this means an infinite redshift: The cosmic redshift, after all, is the ratio of the scale factor at arrival time and the scale factor at emission time, and we have $\lim\limits_{t\to\infty} a(t)/a(t_0) = \infty$. Since the relativistic relative radial velocity of the galaxy on the horizon at the light-emission event and our own galaxy at the event where the light reaches us is related to the Doppler shift, and hence to the scale factor ratio, by eq.~(\ref{CosmologicalDoppler}), it follows from the form of the special-relativistic Doppler factor (\ref{SRDoppler}) that the relative radial velocity must either be equal to the speed of light $c$ in that limit, or else be infinitely large. By construction via parallel transport, the relative radial velocity cannot be larger than $c$. Hence, for galaxies at the cosmological horizon, it must tend to $c$. In consequence, the infinite-time cosmic event horizon is indeed the sphere of galaxies that, if we compare the emission event at $t_0$ and the reception event in the infinitely far future, have radial velocity $c$ relative to our own galaxy. 

\begin{figure*}[htbp]
\begin{center}
% Preprint version
%\includegraphics[width=\textwidth]{figure-conformal-diagram.pdf}
\includegraphics[width=0.75\textwidth]{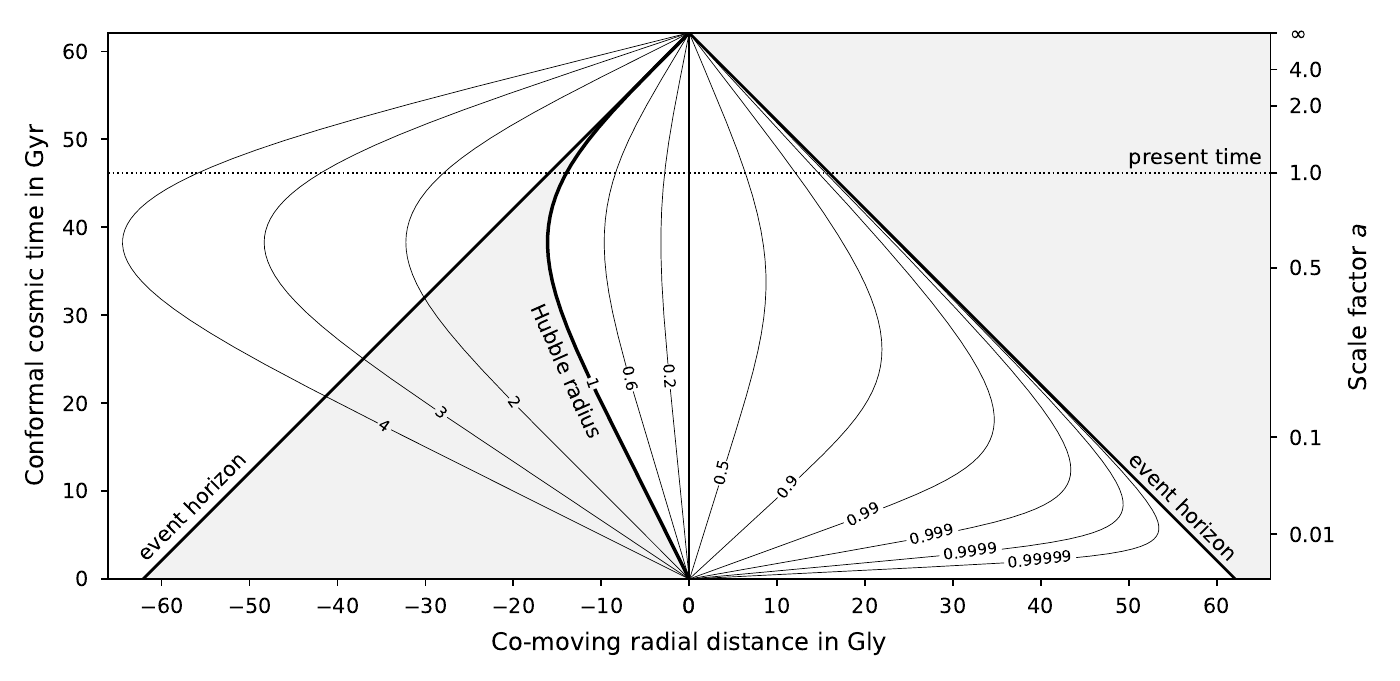}
\caption{Spacetime diagram for conformal time $\tau$ and co-moving cosmic distance $r$,  in units of years and light-years, respectively, for a FLRW universe with $\Omega_{\Lambda}=0.7,\; \Omega_M=0.3,\;H_0=70\;\mbox{km/s}\cdot \mbox{Mpc}^{-1}$. In the left half of the image, contour lines of constant recession speed are shown, expressed as fractions or multiples of the speed of light $c$. In the right half of the image, contour lines of constant relative speed (the magnitude of the radial relative velocity) are shown. Relative speed is between the event at which the value is displayed and the corresponding reception event on our own galaxy's world-line, as in fomula \ref{RadialVelocityCosmo}, with all values given as fractions of $c$
}
\label{Fig:HorizonDoubleFigure}
\end{center}
\end{figure*}

For a model universe that is a good approximation to our own (except for the very early phases of cosmic history), the situation is shown in Fig.~\ref{Fig:HorizonDoubleFigure}. The universe in question is the flat ($K=0$) FLRW model with $\Omega_{\Lambda}=0.7$ and $\Omega_M=0.3$ for the energy density of dark energy and of matter, respectively. To fix the overall scale, I have chosen a Hubble constant value of $H_0=70\;\mbox{km/s}\cdot \mbox{Mpc}^{-1}$. For simplicity, rescale the cosmic scale factor so that $a(t_0)=1$. This allows us to drop the distinction between the radial coordinate $r$ and the present-day co-moving distance $d_{cm}(r,0)$ defined by (\ref{ComovingDefNew}).
In order to deal with the infinite expansion time required for an FLRW model with an infinite-time cosmic event horizon, we introduce the conformal time coordinate $\tau$ defined by
\be
\Dd t = a(\tau)\:\Dd\tau,
\ee
with the Big Bang corresponding to both $t=0$ and $\tau=0$.
In a flat universe, as in our example, and leaving out the angular part, the metric (\ref{FLRWMetric}) becomes
\be
\Dd s^2 = a(\tau)^2(\Dd r-c\:\Dd\tau)(\Dd r+c\:\Dd\tau),
\ee
so for radial light propagation, it follows from $\Dd s^2=0$ that $\Dd r=\pm c\:\Dd\tau$. Choosing appropriate units, such as gigayears for duration and giga-lightyears for spatial distances, light propagation in our spacetime diagram looks just like in special relativity, with light-like geodesics as straight lines, tilted by $45^{\circ}$. From the Friedmann equations that govern the dynamics of expansion, the scale factor as a function of time is given by the implicit equation
\be
\tau-\tau_0 = \frac{1}{H_0}\int\limits^a_1\frac{\Dd a'}{\sqrt{\Omega_Ma'+\Omega_{\lambda}\,a'{}^4}}.
\ee
Fig.~\ref{Fig:HorizonDoubleFigure} shows the cosmic event horizon: the past light-cone of the event on our galaxy's world-line that is infinitely far in the future, corresponding to cosmic time $t=\infty$ and a conformal time of about $\tau=62.1\:\mbox{Gyr}$. We can assign to each point in that spacetime diagram a value for galaxy recession speed, as follows: Each point has a corresponding value for the cosmic time $\tau$ and for the co-moving distance of the galaxy from our own at that time, $r$. That galaxy's proper distance from us at the time corresponds to $d(\tau)=a(\tau)\cdot r$. Given that the Hubble parameter scales as
\be
H(\tau) = H_0\cdot\sqrt{\Omega_M\cdot a(\tau)^{-3}+\Omega_{\Lambda}},
\ee
the Hubble-Lema\^{\i}tre relation yields a recession speed corresponding to
\be
\beta_r = \frac{v_r}{c} = \frac{H_0}{c}\cdot\sqrt{\Omega_M\cdot a(\tau)^{-3}+\Omega_{\Lambda}}\cdot a(\tau)\cdot r.
\ee
Some contour lines for $\beta_r$ are shown in the left half of Fig.~\ref{Fig:HorizonDoubleFigure}. The thicker contour with the value 1, corresponding to $v_r=c$, marks the Hubble sphere. It is clear from the diagram that, apart from asymptotic approximation at very late times, the Hubble sphere is not related to the causal structure of the spacetime shown here --- there is a whole area of galaxies with superluminal recession speeds, shaded grey, whose light is inside the event horizon and thus will still reach us at some future cosmic time.

In the right half of the diagram, we have assigned to each event inside our event horizon a relative radial speed, as follows: If light was emitted from a Hubble flow galaxy at an event at cosmic time $t_e$, that light will reach us at a later time $t_r$. By the given definition, we can compute the relative radial velocity $v_R$ of that galaxy at time $t_e$ to our own galaxy at time $t_r$. The value of that relative velocity is given by equation (\ref{RadialVelocityCosmo}), substituting $t_r$ for $t_0$. Contour lines for $\beta_R\equiv v_R/c$ are shown in the right half of Fig.~\ref{Fig:HorizonDoubleFigure}. As $\beta_R$ gets ever closer to the speed of light,  we see that the corresponding contour line does indeed get closer and closer to forming a boundary that delineates the spacetime regions inside our cosmic event horizon, separating them from the outside regions.

A situation involving this kind of cosmological horizon is analogous to that of a uniformly accelerated observer in special relativity. For each such observer, there will be a light-like ``Rindler horizon.'' When we compare light-like world-lines near that horizon, the transition between light just reaching the observer and the horizon world-line will correspond to $v\to c$ for the observer at the event where the light signal reaches her. A key difference is, of course, that in the special-relativistic situation, we need the observer to be accelerated, while in the presence of gravity, the two world-lines that are diverging in an accelerated fashion can be geodesics. Observers whose world-line is a geodesic do not feel any local gravitational acceleration.

Since four-velocities are defined locally, relating $v\to c$ to a light signal not being able to catch up to a Hubble-flow galaxy is less straightforward than in classical physics. But parallel-transport does allow for statements that are in line with the cannot-catch-up-interpretation. Even in special relativity, a situation where light cannot catch up to a given object can never be formulated in that object's inertial rest frame, where the speed of light is necessarily $c$. It can however, as in the case of the Rindler horizon, be formulated in terms of an external observer charting the positions of both the light signal and the fast-moving object it is catching up with (or not). In our more complicated cosmological scenario, there is a natural external observer, namely the galaxy $G$ whence the light has originated, relative to which our own galaxy is indeed fast-moving. Consider our own galaxy and a distant galaxy $G$ just this side of, but arbitrarily close to, the cosmic event horizon, and consider the event where light from $G$ is just about to arrive at our own galaxy. Using the principle of equivalence, we can describe what happens in the language of special relativity, using the local comoving, and approximately inertial, system defined by our own galaxy's rest frame. In that approximation, the relative acceleration due to gravity is neglected, and what is happening can be described purely in terms of (constant) four-velocities and the distances covered by objects with given four-velocities in a given time interval. Parallel transport of the two four-velocities involved, namely that of the light signal and that of our own galaxy, backwards along the light-like geodesic to the emission event at galaxy $G$, provides an equivalent special-relativistic description, this time in terms of the local standard of rest of $G$ close to the emission event. In that description, our own galaxy is indeed moving away from the light signal, a circumstance that delays the light signal's arrival, and increasing the time the signal needs to cover the last bit of distance. For ever larger values of $v$, it takes longer and longer for the light signal to arrive at our galaxy, with the time needed to cover the last bit of distance diverging for $v\to c$.

While the perspective of the distant galaxy $G$ provides for a description that is compatible with the classical intuition for a situation where one object cannot catch up to another, relativity introduces its own complications. The velocity addition formula (\ref{VelocityAddition}) holds for subluminal speeds, where $v_R=-v_P$ will indeed lead to relative velocity zero, but the universality of the speed of light means that for $v_P=\pm c$, any relative velocity will also be equal to $\pm c$. There is no way of concluding from the statement made using the local standard of rest of $G$ that we, in our own galaxy, will also see that light at finite distance from us will never reach us. The key to the situation is not a matter of speed, but --- as in the case of the Rindler horizon --- involves different measures for length and time intervals applied by different observers. The catch-up picture provides an easily accessible picture for what is happening from the perspective of a distant galaxy near the cosmic event horizon, but not from our own perspective.

\section{Discussion}
\label{Sec:Discussion}
In teaching cosmology at undergraduate or even high school levels, the expanding space interpretation of cosmic expansion is much more common than the relativistic motion interpretation of the cosmological redshift. The expanding space interpretation has the advantage of being closely related to the physical models used for teaching about cosmic expansion at such levels, from one-dimensional rubber bands being stretched via the expanding surface of a rubber balloon being inflated to the three-dimensional model of a raisin cake in the process of being baked, inter-raisin distances changing in a creditable simulation of scale-factor expansion. A proper derivation of the relativistic motion interpretation, on the other hand, requires advanced concepts from the formalism of general relativity (although a simplified toy model exists in the shape of the Milne model [\citealt{Ellis2000}]), which limits its use in introductory teaching.

But as shown in this article, there are definite pedagogical advantages of introducing the relativistic motion interpretation, even without the advanced calculations. Notably, introducing the relativistic radial motion directly via the special-relativistic Doppler formula (\ref{SRDoppler}), allows several useful deductions. One is the absence of superluminal relative speeds of galaxies, eliminating a potential source of confusion for students familiar with special relativity. The second is the Doppler interpretation of the apparent energy loss of redshifted photons travelling from a distant galaxy to our own --- apparently in contradiction to notions of energy conservation if source and receiver are taken to be at rest relative to each other, but readily in line with physical intuition from special relativity in classical physics and special relativity when source and receiver are recognised to be in relative motion. In addition, the interpretation allows students to understand the additional Doppler shifts caused by peculiar motion directly in terms of the special-relativistic velocity addition formula, applied to the radial velocities. Last but not least, as demonstrated in section (\ref{InterpretingCosmicHorizons}), there is an additional pedagogical advantage when it comes to discussing cosmic horizons in infinitely expanding universes. In the relativistic motion interpretation, the existence of such horizons can be understood in terms of the classical intuition of light not being able to catch up to our own galaxy --- given that the relativistic radial motion of our galaxy, as seen from the light-emitting galaxy, approaches the speed of light in the limit that defines the cosmic horizon.

To this author, at least, this set of circumstances suggests that in teaching about the expanding universe at a level that does not introduce the light-propagation formula (\ref{LightSignalIntegral1}), there is a lot to be said for a hybrid approach, which promises to minimise students' propensity to fall prey to common misconceptions about cosmology: Begin in the usual way by teaching about scale-factor expansion, using simple physical models such as the expanding rubber balloon, introducing the Hubble-Lema\^{\i}tre relation and the $v\ll c$ version of the classical Doppler effect to explain the cosmological redshift. At that point, introduce the additional fact that for more distant galaxies, there is a relativistic definition of such galaxies' relative radial velocities, which allows the cosmological redshift to be interpreted using the special-relativistic Doppler formula. On that basis, subluminal motion for distant galaxies, the possible existence of cosmic event horizons and the apparent energy loss of redshifted photons reaching us from other galaxies can all be addressed in a simple fashion that is in line with the arguably most probable set of students' pre-existing conceptions about the relativistic speed limit, energy conservation, and the ability of a signal to catch up, or not, to a fast-moving object.

\section*{Acknowledgements}

I would like to thank Thomas M\"uller for helpful comments on a near-final version of this article, Jorma Louko for pointing me to Schr\"odinger's book on expanding universes, and the anonymous referees for helpful comments and corrections.

\begin{appendix}
\section{Christoffel symbols for FLRW universes}
\label{App:Christoffel}

The metric (\ref{FLRWMetric}) has the non-zero components
\be
g_{00} = -c^2, \;\;\; g_{11} = \frac{a^2(t)}{1-Kr^2}, \;\;\; g_{22} = a^2(t)\:r^2, \;\;\;
g_{33} = a^2(t)\:r^2\:\sin^2\theta.
\ee
Since this metric is diagonal, the components of its inverse are simply
\be
g_{00} = -c^{-2}, \;\;\; g_{11} = \frac{1-Kr^2}{a^2(t)}, \;\;\; g_{22} = \frac{1}{a^2(t)\:r^2}, \;\;\;
g_{33} = \frac{1}{a^2(t)\:r^2\:\sin^2\theta}.
\ee
Recall the definition (\ref{Christoffel2}) for the connection coefficients. It uses the Einstein summation for indices, namely that an identical index symbol that occurs once in an upper and once in a lower position in a specific term, or a product of terms, implies summation, e.g. $A^{\mu}B_{\mu}=A^0B_0+A^1B_1+A^2B_2+A^3B_3$. In the definition (\ref{Christoffel2}) of the Christoffel symbols, that summation applies to the derivatives as well as to the components of the metric and the inverse metric. Directly from the definition, or else with the help of a computer algebra program such as Maxima, \href{http://maxima.sourceforge.net/}{http://maxima.sourceforge.net/}, whose ctensor package provides the functionality for calculating the basic geometric objects from a given metric, one can find that the only non-zero Christoffel symbols for an FLRW spacetime are
\bea
\Gamma^1_{01} &=& \Gamma^2_{02} = \Gamma^3_{03} =\frac{\dot{a}}{a}, \;\;\; \Gamma^0_{11}=\frac{a\dot{a}}{c^2(1-Kr^2)},\;\;\;
\Gamma^{1}_{11} = \frac{Kr}{1-Kr^2},\\[0.5em]
\Gamma^2_{12} &=& \Gamma^3_{13} =\frac{1}{r},\;\;\;\Gamma^0_{22}=\frac{r^2a\dot{a}}{c^2},\;\;\;\Gamma^1_{22}=-(1-Kr^2)r,\;\;\;
\Gamma^3_{23} =\cot\theta,\\[0.5em]
\Gamma^0_{33}&=&\frac{r^2a\dot{a}}{c^2}\sin^2\theta,\;\;\;
\Gamma^1_{33} = -(1-Kr^2)r\:\sin^2\theta,\;\;\;\Gamma^2_{33}=-\sin\theta\:\cos\theta,
\eea
and also any Christoffel symbols that can be obtained from the listed ones by switching the lower two indices (which are symmetric, by the given definition).

Consider (\ref{GeodesicEquation}), noting that since the radial geodesic we are studying is firmly in the $tr$-plane, its $\theta$ and $\phi$ vector components are identically zero. Plugging the relevant non-zero Christoffel symbols into (\ref{GeodesicEquation}) with $\mu=0$, we indeed obtain equation (\ref{TimeGeodesicFLRW}).

\end{appendix}

%\section*{References}

\end{document}